\documentstyle[preprint,prc,aps,epsf]{revtex}
\begin{document}
\draft
\title{Coulomb Distortion Effects for $(e,e'p)$ Reactions at High 
Electron Energy}   
\author{K.S. Kim}
\address{Department of Physics, Yonsei University, Seoul 120-749, 
Korea}
\author{L.E. Wright}
\address{
Institute of Nuclear and Particle Physics,
Department of Physics and Astronomy, Ohio University, Athens,
Ohio 45701}
\maketitle
\begin{abstract}
We report a significant improvement of an approximate method of  
including electron Coulomb distortion in electron induced reactions at
momentum transfers greater than the inverse of the size of the target 
nucleus.  
In particular, we have found a new parametrization for the elastic 
electron scattering phase shifts that works well at all electron 
energies greater than 300 $MeV$.
As an illustration, we apply the improved approximation to the $(e,e'p)$  
reaction from medium and heavy nuclei. 
We use a relativistic ``single particle'' model for $(e,e'p)$ as
as applied to $^{208}Pb(e,e'p)$ and to recently measured data at CEBAF
on $^{16}O(e,e'p)$ to investigate Coulomb distortion effects while
examining the physics of the reaction.
\end{abstract}
\narrowtext

Electron scattering has long been acknowledged as a useful tool for 
investigating nuclear structure and nuclear properties, especially 
in the quasielastic region. 
One of the primary attributes of electron
scattering as usually presented is the fact that in the
electron plane-wave Born approximation, the cross section can be 
written as a sum of terms each with a characteristic
dependence on electron kinematics and containing various bi-linear 
products of the Fourier transform of charge and current matrix elements.  
That is, various structure functions for the process can be extracted 
from the measured data by so-called Rosenbluth separation methods.
The trouble with this picture is that when the Coulomb distortion of 
the electron wavefunctions arising from the static Coulomb field of
the target nucleus is included exactly by partial wave methods,
the structure functions can no longer be extracted from the cross 
section, even in principle.  

Electron Coulomb distortion in elastic and inelastic scattering for 
various processes has been included with various approximations 
in the past\cite{schiff,raven,davies,scheck,duke}.  
In the early 90's Coulomb distortion for the reactions $(e,e')$ and 
$(e,e'p)$ in quasielastic kinematics was treated exactly by the Ohio 
University group\cite{jin1,yates,jin2,zhang,jin3}  using partial wave  
wave expansions of the electron wave functions.  
Such partial wave treatments are referred to as the distorted wave  
Born approximation (DWBA) since the static Coulomb distortion
is included exactly by numerically solving the radial Dirac equation   
containing the Coulomb potential 
for a finite nuclear charge distribution to obtain the distorted  
electron wave functions. 
The induced transition by a virtual photon is included to first order  
(the Born Approximation).  
While this calculation permits the comparison of various nuclear
models to measured cross sections and provides an invaluable check
on various approximate techniques of including Coulomb distortion   
effects, it is numerically challenging and computation time increases   
rapidly with higher incident electron energy.  
And, as noted above, it is not possible to separate the cross section 
into various terms containing the structure functions and develop
insights into the role of various terms in the charge and current 
distributions.

In all of our DWBA investigations of $(e,e')$ and $(e,e'p)$ reactions  
in the quasielastic region, we  used a relativistic treatment based on   
the $\sigma-\omega$ model for the nucleons involved.  
In particular, for the $(e,e'p)$ reaction we use a relativistic 
Hartree single particle model for a bound state\cite{horo}  
and a relativistic optical model for an outgoing proton\cite{hama}  
combined with the free space relativistic current operator
\begin{equation}
J^{\mu} = \gamma^{\mu} + i {\frac {\kappa} {2 M}} 
\sigma^{\mu\nu}\partial_\nu.
\end{equation}
Using this model, we compared our DWBA calculations with experimental  
data measured at various laboratories for $(e,e')$\cite {jin1,yates}, 
and for $(e,e'p)$ \cite{jin2,zhang,jin3} and have found excellent 
agreement with the data.  
We concluded that the relativistic nuclear models are in excellent 
agreement with the measured data and do not need to invoke meson 
exchange effects and other two-body terms in the current that are 
necessary in a Schr\"{o}dinger description that uses a non-relativistic 
reduction of the free current operator \cite{gent}.  
Therefore, in this brief report we will continue to use our relativistic  
``single-particle'' model to investigate Coulomb distortion effects and 
to compare to the newly measured data from CEBAF.

To avoid the numerical difficulties associated with DWBA analyses at 
higher electron energies and to look for a way to still define structure   
functions, our group \cite{kim1,kim2} developed an approximate treatment 
of the Coulomb distortion based on the work of Knoll\cite{knoll} and  
the work of Lenz and Rosenfelder\cite{lenz}.  
Knoll examined approximations of the Green function valid for large 
momentum transfers (that is, valid for $qR>1$ where $R$ is the size of 
the target) while Lenz and Rosenfelder constructed plane-wave-like 
electron wavefunctions which included Coulomb distortion effects. 
We were able to greatly improve some previous attempts along this 
line\cite{giusti,trani} where various additional approximations were 
made which turned out not to be valid.  
We did have the advantage of having the exact DWBA calculation available 
for incident electron energies up to $400-500 MeV$ for checking
our approximations.   
We compared our approximate treatment of Coulomb distortion to the exact  
DWBA results for the reaction $(e,e'p)$ and found good agreement (at  
about the 1-2$\%$ level) near the peaks of cross sections even for heavy  
nuclei such as $^{208}Pb$. The agreement was not so good away from the
peaks.
  
As discussed in our previous papers\cite{kim1,kim2}, one of the ingredients
of our approximate electron wavefunction is a parameterization of the  
elastic scattering phase shifts in terms of the angular momentum.  
In this paper, we briefly review our previous approximation of the 
Coulomb distorted electron wavefunction and present a greatly  
improved parametrization of the phase shifts which works well at all 
incident electron energies greater than 300 $MeV$.  
In addition, we will compare our relativistic 
``single-particle'' model to new experimental data from CEBAF.

Our approximate method of including the static Coulomb distortion 
in the electron wavefunctions is to write the wave functions 
in a plane-wave-like form\cite{kim2};
\begin{equation}
{\Psi}^{\pm}({\bf r})={\frac {p'(r)} {p}}\;e^{{\pm}i{\delta}
({\bf L}^{2})}\;e^{i\Delta}\;e^{i{\bf p}'(r){\cdot}{\bf r}}\;u_{p}\;, 
\label{oldwv} \end{equation}
where the phase factor $\delta({\bf L}^{2})$ is a function of the 
square of the orbital angular momentum operator, $u_{p}$ denotes the  
Dirac spinor, and the local effective momentum ${\bf p}'({\bf r})$ is  
given in terms of the Coulomb potential of the target nucleus by  
\begin{equation}
{\bf p}'({\bf r})=\left( \; p-{\frac{1}{r}} \int^{r}_{0} V(r)dr 
\right ){\bf {\hat p}} \;. \label{lema}  
\end{equation}
We refer to this $r$-dependent momentum as the Local Electron Momentum 
Approximation (LEMA).   
The $ad-hoc$ term $\Delta=a[{\bf {\hat p}}'(r){\cdot}{\hat r}]
{\bf L}^{2}$ denotes a small higher order correction to the electron 
wave number which we have written in terms of the parameter 
$a=-{\alpha}Z(\frac{16 MeV/c}{p})^{2}$. 
The value of 16 MeV/c was determined by comparison with the exact radial 
wave functions in a partial wave expansion.  

The elastic scattering phase shifts are labelled by the Dirac quantum
number $\kappa$ which takes on plus minus integer values beginning with 
one.
The eigenvalues of ${\bf J}^{2}$ are $j(j+1)$ which equals
${\kappa}^{2}-{\frac 1 4}$.  
The basic idea of our approximation is to calculate the elastic  
scattering phases and fit them to function of $\kappa^2$.
Then to replace the discrete values of $\kappa^2$ with the total angular 
momentum operator ${\bf J}^2$ which we then replace by the orbital 
angular momemtum operator ${\bf L}^2$ since the low $\kappa$ terms 
where the difference between $j$ and $l$ is 
significant contribute very little to the cross section.  
The removal of any spin dependence apart from what is in the Dirac 
spinor $u_p$ is crucial for defining
modified structure functions.

Based on earlier work by others we fitted the elastic scattering phases 
shifts to a power series in $\kappa^2$ up to second order;   
\begin{equation}  
\delta_{\kappa}=b_{0}+b_{2}{\kappa}^{2}+b_{4}{\kappa}^{4} \;, 
\label{k2ph} \end{equation}  
where the coefficients, $b_{0}$, $b_{2}$, $b_{4}$ are extracted
from a best fit for $\kappa$ values up to about $3pR$ where $R$ is the 
nuclear radius.  
Note that this procedure requires calculating the elastic scattering
phase shifts for the incident and outgoing electron energies up to 
$\kappa$ values of order $3pR$, which for high electron energies can be 
quite demanding computationally.   
We refer to these phases as the $\kappa^{2}$-dependence phases. 
This fit to the phases worked very well for $\kappa$ values up to   
approximately $\kappa=3pR{\approx}35$ at medium or low energy, but did 
not fit the exact phases shifts very well for higher energies where 
$\kappa=3pR{\ge}50$ or more. 
Since we were primarily looking at electron energies in the 300-600 MeV 
range in our previous work, this discrepancy did not present 
a significant problem. 

However, with CEBAF type energies we need a fit to the phases that will 
work at any incident energy where the overall approximation can be used;  
that is, for incident electron energies greater than about 300 $MeV$ and 
processes with momentum transfers greater than about $1/R$.  
In addition, we would like to avoid calculating all of the elastic 
phase shifts, particularly the very high ones. 
A reasonable solution is to make use of the fact that the higher  
$\kappa$ phase shifts appoach the point Coulomb phases which have  
a simple analytical form at high energy.  
At the other extreme, the low $\kappa$ phases corresponding to orbitals 
which penetrate the nucleus are linear in $\kappa^2$ which was the basis 
or our initial parametrization. 
The difficult phases to fit correspond to $\kappa$ values of order  
$pR$ which, from a classical point of view, corresponds to scattering
from the nuclear surface.  
Moreover, it is well known that in electron induced reactions the spatial
region around the surface gives the largest contribution to the 
cross section, so it is important to fit  
the intermediate range as well as possible.

\begin{figure}[p]
\newbox\figk
\setbox\figk=\hbox{
\epsfysize=150mm
\epsfxsize=150mm
\epsffile{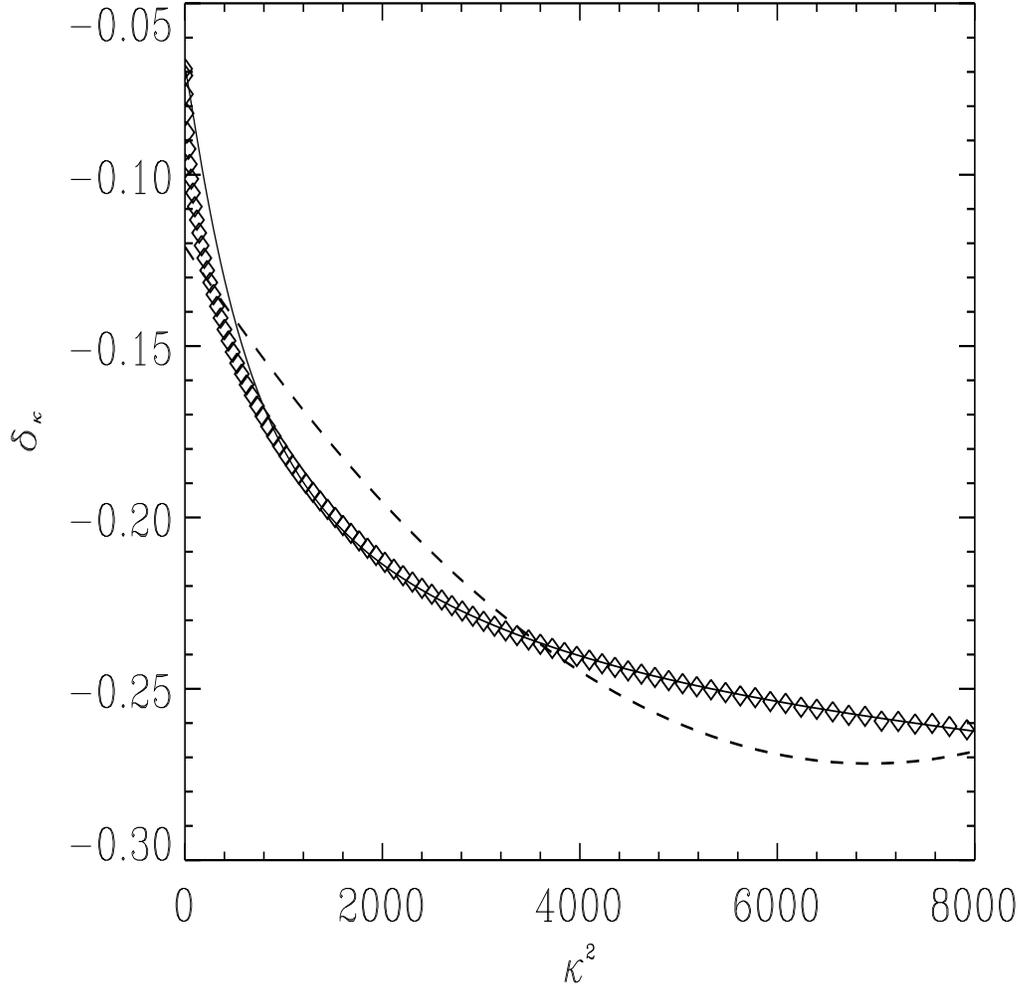}}
\noindent\hspace{10mm}\vspace{5mm}\box\figk
\caption[phop]{The comparison between the exact, $\kappa^{2}$-dependent,  
and new phases in $^{16}O$ for $\kappa_{max}=100$ and energy $E=2441$ 
MeV. The diamonds are the exact phases, the dashed curve for $\kappa^{2}$
fit, and the solid curve for the new phase shift parametrization.}  
\label{phop}
\end{figure}

Another goal is to reduce the computer time needed, so we decided to seek
a parametrization of the elastic scattering phases shifts in terms of 
$\kappa^{2}$ which has the correct large $\kappa^2$ behaviour and becomes 
linear in $\kappa^2$ at low angular momentum.
Since we have the correct large $\kappa$ behaviour, we need only  
calculate the exact scattering phase shifts for $\kappa$ values of
up to of order $pr$.  
The large $\kappa$ and small $\kappa$ behaviour are quite
different, so we chose to write the expression for the phase shift as the 
sum of two terms with an exponential factor which suppresses one of the 
terms at small $\kappa$ values and the other at large $\kappa$ values. 
After some experimentation, we find that the following parametrization 
of elastic scattering phase shift describes the exact phase shifts 
very well:
\begin{eqnarray}
{\delta}({\kappa})&=&[a_0 + a_2\frac{{\kappa}^{2}}{(pR)^{2}}]
e^{-{\frac{1.4{\kappa}^{2}}{(pR)^{2}}}} 
\nonumber \\ 
&&-{\frac{{\alpha}Z}{2}}(1-e^{-{\frac{{\kappa}^{2}}{(pR)^{2}}}}) 
{\times}{\ln}(1+{\kappa}^{2})  \;,  \label{newph} 
\end{eqnarray}
where $p$ is the electron momentum and we take the nuclear radius to be
given by $R=1.2A^{1/3}-0.86/A^{1/3}$.  We fit the two constants $a_0$ and
$a_2$ to two
of the elastic scattering phase shifts ($\kappa=1$ and $\kappa=Int(pR)+5$).
To a very good approximation, $a_0=\delta(1)$ and
$a_2=4\delta(Int(pR)+5)+\alpha Z ln(2pR)$, where $Int(pR)$
replaces $pR$ by the integer just less than $pR$. 
Note that this parametrization only requires the value of the exact 
scattering phase shift for $\kappa=1$ and $\kappa=Int(pR)+5$.
As shown in Fig. \ref{phop}, the ${\kappa}^{2}$-dependence phase 
parametrization breaks down for high $\kappa$ values and has large 
deviations for mid-range $\kappa$ values.  
The new phase parametrization fits the exact phases very well for 
electron energy of $E=2400$ MeV on $^{16}O$, although  
the new phase parametrization does still show some small deviations
from the exact phases for $\kappa$ values around $20$ to $30$ which is 
in the surface region.  
Clearly additional terms could be added to the parametrization
to obatin a better fit.  
However, as we shall see below, the simple fit that we have used 
reproduces the cross section quite well.  

Using the new phase shift parametrization and the local effective 
momentum approximation we construct plane-wave-like wave functions 
for the incoming and outgoing electrons.  
Since the only spinor dependence is in the Dirac spinor all of the 
Dirac algebra goes through as usual and we end up with a M{\o}ller-like 
potential which contains an $r$-dependent momentum transfer.  
It is then straightforward to calculate the $(e,e'p)$
cross sections and modified structure functions. 
Please see our previous papers \cite{kim1,kim2} for details. 

In most $(e,e'p)$ experiments, there is sufficient energy resolution 
that protons knocked out of different shells can be examined.  
It is common to report the experimental results in terms of the reduced  
cross section $\rho_{m}$ as s function of missing momentum $p_{m}$, 
which is defined by
\begin{equation}
{\rho}_{m}(p_{m})={\frac{1}{PE_{p}{\sigma}_{eP}}}
{\frac{d^{3}{\sigma}}{dE_{f}d{\Omega}_{f}d{\Omega}_{P}}} \;, 
\label{rdcs} \end{equation}
where the missing momentum is determined by the kinematics
${\bf p}_{m}={\bf P}-{\bf q}$ where ${\bf P}$ is the outgoing 
proton momentum and ${\bf q}$ is the asymptotic
momentum transfer from the electron defined by 
${\bf q}={\bf p}_{i}-{\bf p}_{f}$.  
For plane wave protons  in the final state $\rho_{m}$  is 
related to the probability that a bound proton from a given shell  
have momentum ${\bf p}_{m}$.  
For the off-shell electron-proton cross section ${\sigma}_{eP}$
we use  the form `cc1' given by de Forest \cite{defo}.  
For distorted outgoing protons, this reduced cross section is just 
a convenient way of comparing experiment and theory since the theory 
results for the cross section can have the same factors removed. 
Note that all calculations will be carried out 
in the laboratory system (target fixed frame). 

While there are two experimental kinematic arrangements commonly used 
in $(e,e'p)$ experiments with designations of
parallel kinematics and perpendicular kinematics, in the present work, 
we consider only  perpendicular kinematics.  
In perpendicular kinematics, the momentum transfer ${\bf q}$ is held 
fixed along with the magnitude of the momentum of the outgoing proton 
while the angle between ${\bf q}$ and ${\bf P}$ is varied.  
The calculated reduced cross section is compared (by means of a
linear least squares fit) to the similarly 
reduced experimental cross section to extract an overall scale
factor which is the spectroscopic factor.  The spectroscopic factor 
contains two factors, the occupation probability of a proton
in a given orbit and the overlap of the residual nucleus with the 
$A-1$ nucleons in the target.

\begin{figure}[p]
\newbox\figk
\setbox\figk=\hbox{
\epsfysize=150mm
\epsfxsize=150mm
\epsffile{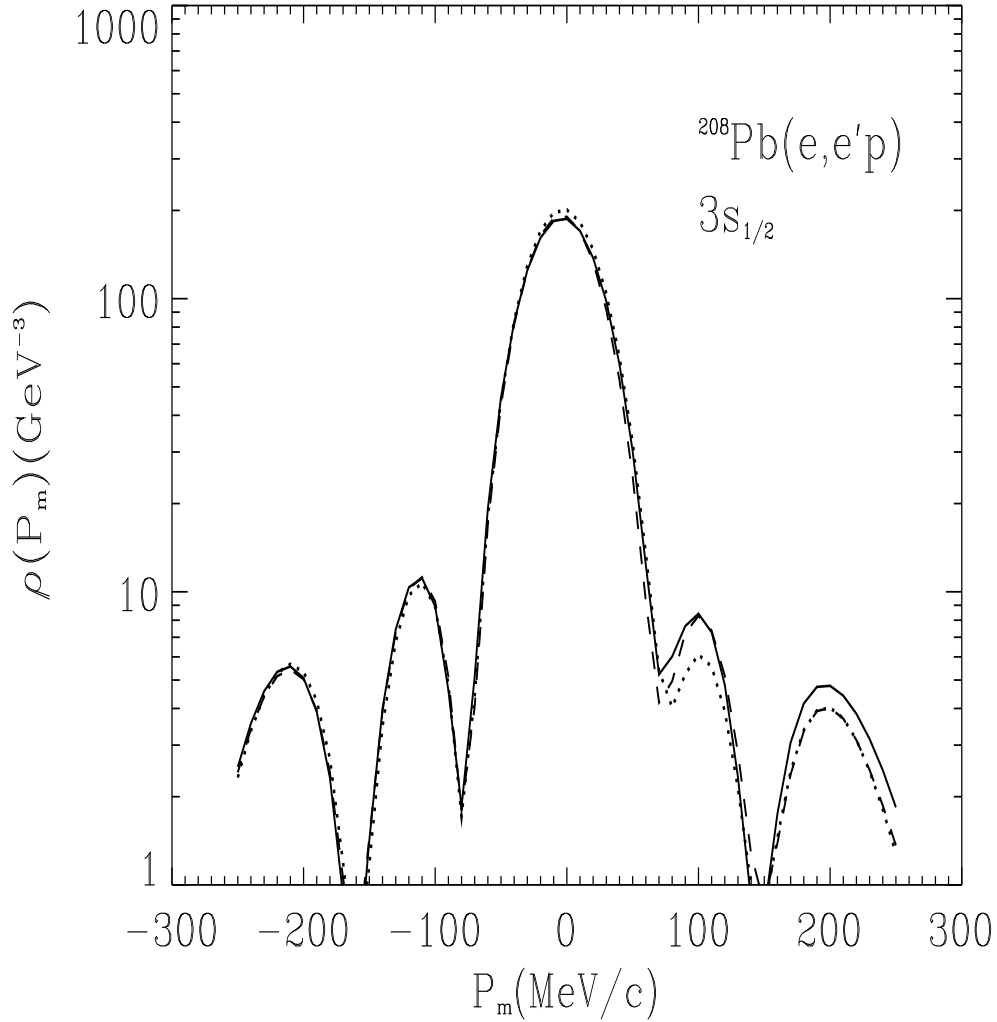}}
\noindent\hspace{10mm}\vspace{5mm}\box\figk
\caption[pbs]{The reduced cross section for $^{208}Pb$ for the $3s_{1/2}$ 
orbit with perpendicular kinematics.  The kinematics are $E_{i}=412$  
MeV, and proton kinetic energy $T_{P}=100$ MeV.  The solid line is the 
full DWBA result, the dashed line is the approximate DWBA using 
the new phase shift parametrization, and the dotted line is the approximate 
DWBA with the $\kappa^{2}$-dependent phases.}  
\label{pbs}
\end{figure}

As a test case, we calculate the reduced cross sections with the new 
phases for a heavy nucleus, $^{208}Pb$. 
Figure \ref{pbs} shows the reduced cross section as a function of the 
missing momentum $p_{m}$ for knocking protons from the $3s_{1/2}$ 
orbital of $^{208}Pb$.  
The incident electron energy $E_{i}=412$ MeV, and
the outgoing proton kinetic energy is $T_{p}=100$ MeV.  
We have chosen $P=q$ which corresponds to an electron scattering angle 
of $\theta_e=74^o$.
The solid line is the result of the full DWBA \cite{jin2}, the dashed  
curve is the result with the new phase shift parametrization, and the 
dotted curve is the result with $\kappa^{2}$-dependence phase shift 
parametrization. 
The dashed curve obtained by using the new phases clearly reproduces 
the exact result much better than the previous ${\kappa}^{2}$-dependence 
phase parametrization over the whole region.   

\begin{figure}[p]
\newbox\figk
\setbox\figk=\hbox{
\epsfysize=150mm
\epsfxsize=150mm
\epsffile{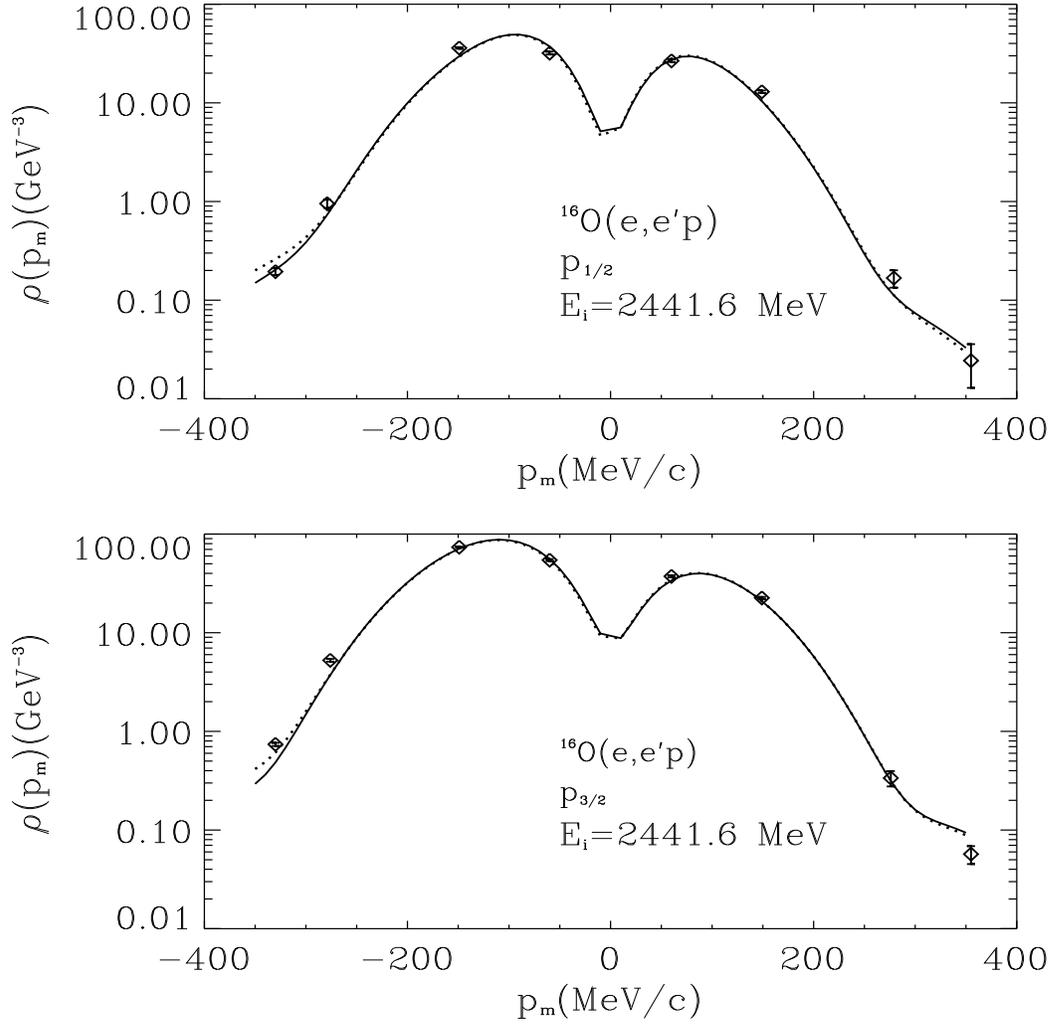}}
\noindent\hspace{10mm}\vspace{5mm}\box\figk
\caption[rd244]{The cross sections for $^{16}O$ from $p_{1/2}$ and 
$p_{3/2}$ orbits for perpendicular kinematics.  The kinematics 
are $E_{i}=2441.6$ MeV, proton kinetic energy $T_{P}=427$ MeV, and 
energy transfer $\omega=436$ MeV.  The solid lines are the new 
approximate DWBA results, the dotted lines are the PWBA results, and 
the data are from CEBAF \cite{gao}.} 
\label{rd244}
\end{figure}

We also apply the new phase shift paramtetrization for the case of high 
energy electron scattering on the light nucleus $^{16}O$ where protons 
are removed from the $p_{1/2}$ and $p_{3/2}$ orbits. 
The incident electron energy $E_{i}=2441.6$ MeV and the outgoing proton
kinetic energy  $T_{p}=427$ MeV as shown Fig. \ref{rd244}.  
In this figure, the solid curves are the approximate DWBA results using
the new phase shift parametrization, the dotted curves are the PWBA results
without Coulomb distortion, and the data are newly measured from 
CEBAF as reported in the dissertation of Gao \cite{gao}.  
Note that our exact DWBA code cannot evaluate such high energy processes 
without extensive modification which we have not done. 

As expected, the effect of Coulomb distortion on such a high energy 
electron induced process is very small except possibly at large missing
momentum.  
Note that the Coulomb effects for $^{16}O$  in the medium energy 
region (500 MeV) was of the order of $3\%$ \cite{jin3}. 
This fit to the experimental data using our relativistic ``single particle''
model for the nucleon wavefunctions results in spectroscopic factors of
61$\%$ for the $p_{1/2}$ orbital and 70$\%$ for the $p_{3/2}$ orbital.
In our analyis of Saclay data \cite{chin} at lower electron energies
using a similar nuclear model we found spectroscopic factors
of 54\% and 57\% respectively \cite{jin3}.  

In summary, we have improved our previous approximate method of
including Coulomb distortion effecs in $(e,e'p)$ reactions from
nuclei.  The improvement involves a better parametrization of
the elastic scattering phases shifts which has the correct behaviour for
large angular momenta and requires the calcuation of only two phase shifts (for
$\kappa =1$, and for $\kappa$ equal to $Int(pR)+5$).  We showed that
even for $(e,e'p)$ on $^{208}Pb$ the cross section calculated with
our approximation using the improved
parametrization of the phase shifts agrees with the exact DWBA result quite well even
out beyond the second maxima.  This is a significant improvement over
our previous approximation for the phase shifts.  In addition, we
compared our relativistic ``single-particle'' model for $(e,e'p)$ from
$^{16}O$ to the recently measured cross section at Thomas Jefferson Lab and found
excellent agreement for the removal of a proton from the $p_{3/2}$ and
$p_{1/2}$ shells with reasonable spectroscopic factors.

Our improved approximate method of including Coulomb
distortion in electron scattering reactions  works for high energy
electrons as well as for more moderate energies ($300-500 MeV$), and for experiments
at the few percent level this approximate way of including Coulomb distortion
is adequate.  More importantly, as discussed in our previous
paper, this ``plane-wave-like'' approximation permits the
extraction of ``structure functions'' even in the
presence of strong Coulomb effects and thus provides a very good tool
for looking into the response of the nucleus to ``longitudinal'' and
``transverse'' photons.

\end{document}